# RNN-Based Models for Predicting Seizure Onset in Epileptic Patients

Mathan Kumar Mounagurusamy
Department of Computing Technologies
School of Computing
SRM Institute of Science and
Technology, Kattankulathur,
Chennai, Tamil Nadu, India
drmathan16@gmail.com

Thiyagarajan V S
Department of CSE
Karpaga Vinayaga College of
Engineering and Technology,
Chengalpattu, Tamil Nadu, India
thiyagu.cse86@gmail.com

Abdur Rahman
Master of Science in Information
Technology,
Department of Accounting and
Technology, School of Business &
Technology,
Emporia State University, United States
of America
arahman.esu@gmail.com

Shravan Chandak
Symbiosis Centre for Management
Studies, Pune,
Symbiosis International (Deemed
University), Pune, India
shravan.chandak@scmspune.ac.in

D. Balaji
Department of Electronics and
Communication Engineering
Vel Tech Rangarajan Dr. Sagunthala
R&D Institute of Science and
Technology,
Chennai, Tamil Nadu, India
dbalaji@veltech.edu.in

Venkateswara Rao Jallepalli
Department of Pharmacy Practice
MB School of Pharmaceutical Sciences
Mohan Babu University
Tirupati, Andhra Pradesh, India
venkateswararao.j92@gmail.com

*Abstract--* **Early management and better clinical outcomes for epileptic patients depend on seizure prediction. The accuracy and false alarm rates of existing systems are often compromised by their dependence on static thresholds and basic Electroencephalogram (EEG) properties. A novel Recurrent Neural Network (RNN)-based method for seizure start prediction is proposed in the article to overcome these limitations. As opposed to conventional techniques, the proposed system makes use of Long Short-Term Memory (LSTM) networks to extract temporal correlations from unprocessed EEG data. It enables the system to adapt dynamically to the unique EEG patterns of each patient, improving prediction accuracy. The methodology of the system comprises thorough data collecting, preprocessing, and LSTM-based feature extraction. Annotated EEG datasets are then used for model training and validation. Results show a considerable reduction in false alarm rates (average of 6.8%) and an improvement in prediction accuracy (90.2% sensitivity, 88.9% specificity, and AUC-ROC of 93). Additionally, computational efficiency is significantly higher than that of existing systems (12 ms processing time, 45 MB memory consumption). About improving seizure prediction reliability, these results demonstrate the effectiveness of the proposed RNN-based strategy, opening up possibilities for its practical application to improve epilepsy treatment.**

*Keywords: Seizure Prediction, Epilepsy, Electroencephalogram, Clinical Decision Support Systems, Prediction Accuracy, False Alarms*

I. INTRODUCTION

Millions of people worldwide suffer from epilepsy, and because it is unpredictable, managing patients with it can be extremely difficult. To enhance therapeutic results and patient quality of life, seizure prediction algorithms have come to light as potentially useful early intervention tools. Static thresholds and simple EEG features are frequently used in traditional methods, which results in inadequate precision and high false alarm rates. Novel approaches that can capture the intricate temporal dynamics included in EEG signals are needed to overcome these limitations. The pressing requirement to increase seizure prediction accuracy while reducing false alarms is what spurred the research. More complex models are required since existing systems frequently cannot adjust to the variety of individual patients and dynamic EEG patterns. The study provides a novel method to greatly increase the accuracy and reliability of seizure start prediction by utilizing developments in deep learning, namely RNNs fitted with LSTM cells. Creating an RNN-based system that can predict the beginning of seizures in epileptic patients more accurately than current approaches is the main goal of the study. The method aims to extract complex temporal correlations directly from unprocessed EEG data by utilizing LSTM networks. The technique improves prediction accuracy while lowering false alarms, which allows for prompt clinical interventions and raises the standard of care for patients as a whole. The study uses LSTM networks to directly extract complicated temporal correlations from raw EEG data, thereby contributing a novel RNN-based methodology for seizure onset prediction in epilepsy. The research contributes to the development of more dependable clinical decision support systems by greatly increasing prediction accuracy, lowering false alarm rates, and increasing computing efficiency. Better management and treatment outcomes for epilepsy are anticipated thanks to the suggested framework, which also improves the understanding of the EEG dynamics before seizures and provides a flexible and scalable method for individualized patient care. The proposed RNN-based system for seizure prediction in epileptic patients is carefully explored and presented in the study, which is divided into several important components. Section I provides an overview of the study's background and motivations, emphasizing the limitations of current approaches and the demand for advanced predictive models. Section II examines relevant research in the realm of seizure prediction and talks about different strategies and their drawbacks. The methodology of the proposed system is expounded upon in Section III, encompassing data collection, preprocessing, LSTM-based feature extraction, model design, and optimization techniques.

In Section IV, the analysis and empirical results are presented. The performance of the proposed system is compared with existing systems in terms of computational efficiency, accuracy, and false alarm rates. The implications of the results, possible disadvantages, and ideas for future research are discussed in Section V. In summary, Section VI brings the study to a conclusion highlighting the contributions of the research and highlighting how RNN-based models can improve patient care and epilepsy management.

## II. RELATED WORK

Epileptic seizures are the means by which the networks of an epileptic individual grow throughout time and place. The study develops a generalizable approach to forecast a specific patient seizure by evaluating feature representation and extracting features from multichannel EEG data. Using the supplied parameters, the peculiarities of the EEG signals are exposed. The features are sent into the r - ELM), which uses them to assess feature representation and train the data as a whole [6]. The prediction of epileptic seizures using EEG data is the main topic of the present study. Unlike the conventional technique, which assumes that the preictal state lasts the same amount of time throughout each of a patient's seizures, it suggests a novel strategy that uses clustering to identify each seizure separately. Actually, it increases the efficiency of the binary classifier that distinguishes between the interictal and preictal stages and lowers label noise [7]. In order to produce synthetic EEG data, the paper suggests using a DCGAN. It also aims to evaluate four popular DL algorithms' ability to predict epileptic seizures using TL. Methods: Using DCGAN trained on actual EEG data in a patient-specific way, it proposed a technique that generates synthetic data. It uses CESP, a novel approach, and one-class SVM to assess the quality of the produced data [8]. It introduces a unique deep learning system that uses multi-channel EEG observations from human scalps to predict epileptic seizures. Based on the categorization of the epileptic patient's brain states into interictal and preictal categories, a patient-specific system is suggested. For autonomous feature learning and classification, the system trains a dual-dimensional convolutional variational autoencoder once under supervision [9]. A seizure prediction system's goal is to accurately pinpoint the pre-ictal brain state, which comes before to a seizure incident. Seizures may be predicted accurately for several subjects in a dataset using patient-independent seizure prediction models, which have been shown to be a practical solution to the issue [10]. In the present study, it offers hardware-friendly and energy-efficient approaches for seizure prediction in epilepsy. The neural architecture hunt produced a model that weighed just forty-five kB, which was assessed on three different datasets [11]. It proposes and evaluates two deep learning models: one based on a CNN and the other on GRU-LSTM. The models are used to forecast when a seizure will start by differentiating between the preictal and interictal states. utilizing five electrodes for EEG readings and limiting the warning time to only ten minutes before to the start of the seizure, the procedures were centered on the comfort of the patient while utilizing the device [12]. The purpose of the project was to introduce three ML techniques that use the characteristics of electrocardiogram data to predict epileptic episodes. The ECG data collection, which included thirteen individuals, was first pre-processed. The results demonstrate that the Naive Bayes model is a good option for forecasting epileptic episodes [13]. The proposed approach begins by teaching a STN to recognize temporally and intensity invariances in EEG data. In order to provide input characteristics for a CNN, the proposed method also computes the modified EEG signals' STFT. One technique to lessen the relevance of localized false positives is to use a k-out-of-n post-processing procedure [14]. In the research, the DTF and CNN were combined to propose a new approach for patient-specific forecasting of seizures by taking into account the unique information flow across EEG channels from the viewpoint of entire brain activity. First, the DTF method was used to segment the iEEG impulses and quantify their information flow properties. Next, based on channel pairings and information flow frequency, these characteristics were recreated as channel-frequency maps [15]. In the present investigation, the authors employ time-frequency extraction and classification of features approaches to anticipate a seizure with excellent prediction accuracy beforehand from EEG data. Clinicians and caregivers will be better able to appropriately intervene via medication or other preventative measures if these can anticipate an epileptic seizure early on [16].

## III. PROPOSED SYSTEM

To address the shortcomings of the existing system, provide a novel RNN based system in the study for seizure onset prediction in epileptic patients. Existing systems frequently rely on static thresholds and basic features obtained from EEG data, which might result in a higher rate of false alarms and less accurate predictions. These limitations highlight the pressing requirement for more advanced models that can accurately represent the intricate temporal dynamics seen in EEG signals. The proposed system is very different from the existing system in many important areas. First, it uses RNNs to make use of the temporal dependencies that are naturally present in EEG data, rather than depending on manually created features or fixed thresholds. RNNs are useful for capturing temporal patterns and long-term dependencies in sequential data, such as EEG signals. It raises the precision and resilience of seizure prediction by allowing the model to dynamically adjust to variations in EEG patterns across time. Operational Flowchart is shown in fig.1. The proposed system must be put into place in a few different steps. EEG data is first preprocessed to normalize the signals and extract pertinent information. In contrast with traditional techniques, which frequently depend on made characteristics like amplitude thresholds and spectral power, the proposed system feeds the RNN with raw EEG data straight from the source. The technique lowers the risk of information loss associated with handmade features by enabling the model to automatically learn and extract useful features from the data. After that, a specific RNN architecture created to capture the temporal dependencies between subsequent EEG samples is fed the preprocessed EEG data. In particular, it uses an RNN version called the LSTM network, which is renowned for its capacity to learn and retain long-term dependencies. In comparison to conventional techniques, the model's ability to efficiently learn the complex patterns preceding seizure onset is improved by the LSTM design. The RNN-based model gains knowledge from previous EEG data that has been associated with seizure episodes during training. The model optimizes its capacity to precisely identify seizure beginning and reduce prediction errors by iteratively adjusting its parameters. It uses rigorous validation methods, like cross-validation and testing on separate datasets, to make sure the model is reliable and generalizable to a variety of patient profiles and EEG recording scenarios.

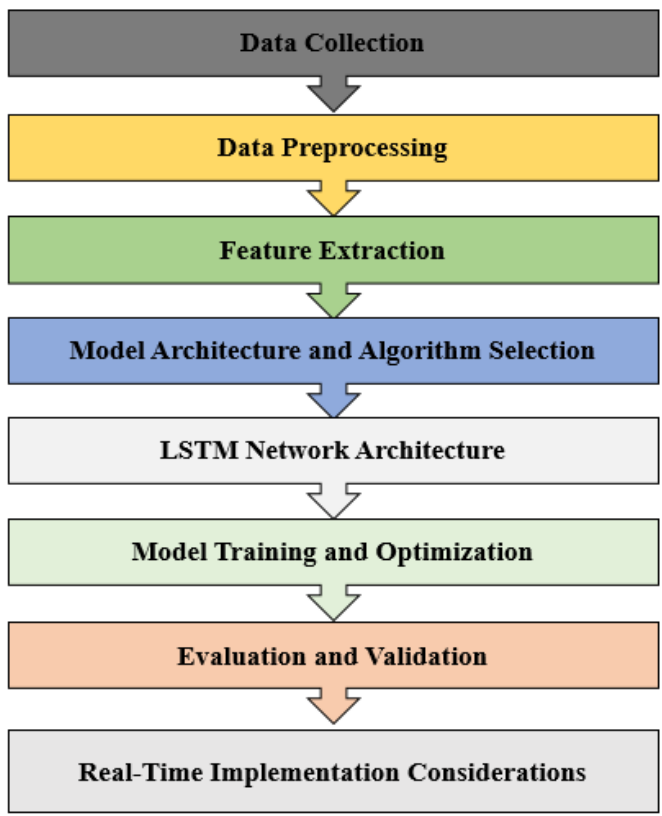

Fig.1. Operational Flowchart

There are numerous benefits to putting the proposed system into practice. First off, using RNNs can improve seizure start detection's predictive power, which could lead to early intervention and better patient care. Because RNNs are dynamic, the model can adjust to the unique characteristics of each patient as well as changes in EEG patterns over time, producing more accurate and customized prediction results. Furthermore, by using an automated feature extraction technique, the proposed system becomes more scalable and adaptable to a variety of clinical contexts by reducing the need for human participation and domain-specific knowledge. In summary, the proposed RNN-based system provides a major improvement over the existing system for epilepsy seizure prediction. The objective is to give clinicians a more accurate and dependable tool for forecasting the start of seizures by utilizing deep learning and temporal modeling. It will ultimately improve patient outcomes and quality of life. To improve the model's therapeutic usefulness, future work will concentrate on enhancing its design, investigating new EEG modalities, and adding real-time monitoring features.

*A. Data Collection:*

EEG recordings from epileptic patients undergoing clinical monitoring are collected as part of the data collection for the study. The procedure carefully complies with ethical standards, which include getting participants' and their guardians' informed consent. Multiple sessions of EEG recordings are made to capture a wide variety of seizure types and cover a wide range of patient demographics. By taking these steps, the training and assessment datasets are guaranteed to be representative and thorough, which improves the generalizability of the generated seizure prediction model. Patients are observed using established clinical protocols during the collection of EEG data; depending on the clinical demands and required period of monitoring, it may involve both inpatient and outpatient settings. By the worldwide 10-20 system, EEG electrodes are positioned on the scalp to efficiently record brain activity across several regions. To record both ictal (seizure) and interictal (non-seizure) events are naturally occur, continuous recording is usually done. To ensure data quality and consistency between sessions and patients, particular attention is paid to recording factors such as sampling rate, electrode location, and session duration. Seizures diaries and medical histories are examples of supplementary clinical data that are collected to give context for EEG interpretation and model building.

*B. Data Preprocessing:*

Preprocessing EEG data is necessary to ensure its quality and appropriateness for further analysis with RNNs. To prevent noise and artifacts from distorting the underlying brain signals, the data is first carefully filtered. It is an important step in reducing interference and enhancing signal clarity, both of which potentially compromise seizure prediction accuracy. Normalization is used to equalize signal amplitudes across all recordings after filtering, guaranteeing consistency in data representation and enabling comparison between various patients and sessions. The preprocessed EEG data is then divided into three epochs that represent different times: pre-ictal (before a seizure), ictal (during a seizure), and inter-ictal (between seizures). By segmenting the data, the RNN model can identify distinct temporal patterns linked to the start of seizures and differentiate them from baseline brain activity. These preprocessing procedures methodically prepare the EEG data, reducing variability and optimizing model performance to provide a strong basis for precise and dependable seizure prediction in clinical applications.

*C. Feature Extraction:*

The methodology uses feature extraction from EEG data and deviates from conventional methods by taking advantage of LSTM network capabilities. Instead of using hand-crafted features like amplitude thresholds or spectral power use raw EEG data sequences as direct inputs to the LSTM network. LSTM networks are chosen in particular because of their ability to capture long-term dependencies in sequential data, which is an essential feature for identifying subtle patterns in epilepsy that occur before seizures. The technique reduces information loss and improves forecast accuracy by allowing the model to learn and extract relevant characteristics directly from the EEG data. The methodology not only improves the fidelity and reliability of seizure event detection and prediction in the model but also maximizes the model's ability to harness the temporal dynamics inherent in EEG data using LSTM networks. An important development in seizure prediction techniques is the move towards automated feature extraction inside a deep learning framework, allowing for more sophisticated insights and better clinical outcomes in the treatment of epilepsy.

*D. Model Architecture and Algorithm Selection:*

The proposed technique uses a specialized RNN architecture, namely Long Short-Term Memory (LSTM) networks, to successfully capture the subtle temporal relationships in EEG data that are crucial for forecasting seizure onset. Unlike classic approaches that use static thresholds or created features, the LSTM network may dynamically learn and preserve long-term relationships from raw EEG data. It is intended to overcome the vanishing gradient issue that exists in traditional RNNs, making it appropriate for learning from series with long-range dependencies. The model is made up of numerous stacked LSTM layers that support hierarchical feature extraction, allowing the network to gradually learn more abstract and complicated patterns linked to seizure onset. The LSTM architecture was chosen over traditional models for its improved capacity to simulate the temporal fluctuations of EEG data, resulting in more accurate and dependable seizure predictions.

*E. LSTM Network Architecture:*

The methodology's main component is the application of a customized LSTM network made just for seizure prediction. Since LSTM networks can solve the vanishing gradient issue that typical RNNs have, it makes them especially suitable for learning from sequences with long-range dependencies, which are common in EEG data processing for epilepsy prediction. A memory state that can selectively keep and delete information over time steps is a key component of the distinctive cell structure of the LSTM architecture. The capacity is essential for recognizing and analyzing the intricate temporal patterns present in EEG signals, which are necessary for identifying the minute alterations that occur before the beginning of seizures. Input, forget, and output gates are the gates that control the information flow in the network and are how the LSTM cell functions. These gates allow the LSTM to efficiently learn both short-term and long-term dependencies by allowing it to maintain and update its memory state in response to the sequential input data. It enables hierarchical learning of features at various levels of abstraction in the implementation by stacking numerous layers of LSTM cells. The network's capacity to distinguish between pre-seizure patterns and normal brain activity is improved by its hierarchical structure, which enables it to extract progressively complicated information from the raw EEG data. Sequences of preprocessed EEG data that have been labeled with matching seizure onset events are fed into the LSTM network during training. Using backpropagation, the network iteratively modifies its internal parameters to maximize its capacity to precisely forecast seizure episodes based on learned patterns and past data. To improve the network's forecast accuracy across subsequent epochs, the training method entails minimizing a loss function that measures the difference between predicted and actual seizure timings. Logic Diagram is shown in fig.2.

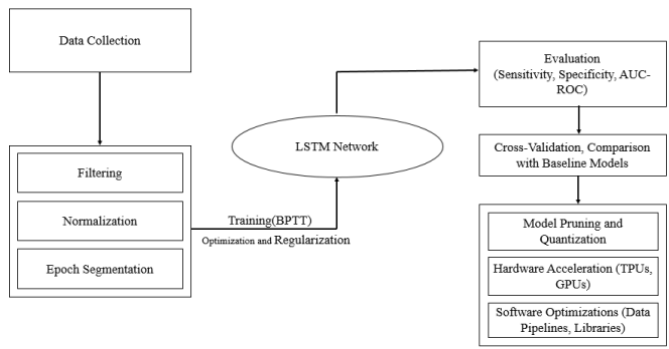

Fig.2. Proposed Logic Diagram

*F. Model Training and Optimization:*

Annotated EEG data, in which the timing of seizure onset instances is explicitly recorded, is used to train the LSTM network. For supervised learning to work and teach the model the connection between EEG patterns and later seizures, labeled data is a must. Feeding the LSTM network with batches of previously processed EEG data is the first step in the training process. Every sequence contains both times when the brain is not having seizures and times when it is. Backpropagation through time (BPTT) is the technique by which the model's internal parameters are iteratively adjusted during training. The backpropagation algorithm's BPTT modification was created to manage the sequential nature of time-series data, such as EEG. For every sequence, the LSTM network is unrolled through time. The gradients of the loss function for each parameter are then computed, and the weights are updated using the gradients. Typically, the loss function guides the optimization process by measuring the discrepancy between the actual seizure occurrence and the projected probability of one. The model improves its forecast accuracy and detects tiny pre-seizure patterns by fine-tuning its internal weights with each iteration. To ensure that the model generalizes adequately to new, unknown data and avoids overfitting, regularization techniques like dropout can be used. The model learns to correlate complex EEG data to precise seizure predictions by methodically refining the LSTM network through BPTT. It produces a dependable tool for early detection and intervention in clinical situations.

*G. Evaluation and Validation:*

It utilizes rigorous assessment criteria, such as sensitivity, specificity, and the area under the receiver operating characteristic curve (AUC-ROC), to thoroughly evaluate the efficacy of the LSTM-based seizure prediction model. Specificity assesses the model's precision in identifying non-seizure periods, whereas sensitivity gauges the model's capacity to accurately recognize real seizure episodes. When comparing the model's overall discriminative power across different threshold settings, the AUC-ROC provides a balanced metric. Cross-validation procedures are used to validate the prediction power of the model. It entails dividing the dataset into several folds and training and testing the model on various subsets iteratively. Cross-validation improves the model's generalizability by ensuring that its performance is stable and not unduly reliant on any specific patient dataset and EEG recording setting. To demonstrate the benefits of the LSTM approach also contrast the efficacy of the model with baseline techniques, such as conventional threshold-based or manually constructed feature-based models. To establish a standard for the clinical relevance and effectiveness of the model, expert annotations from neurologists are compared with the predictions of the model. To confirm the accuracy and dependability of the seizure prediction model and ensure its successful integration into clinical practice to promote early intervention and enhance patient outcomes, these thorough evaluation and validation procedures are essential.

*H. Real-Time Implementation Considerations:*

The methodology takes into account several important factors for clinical deployment to ensure the effectiveness and efficiency of the LSTM network's real-time prediction capabilities. It is essential to maximize computing effectiveness and reduce latency without compromising prediction accuracy. To do that investigate methods like model pruning, which shrinks the size of the network by getting rid of unnecessary and unimportant parameters, lowering the computing burden, and accelerating the inference process. Another important technique is quantization, which preserves model performance while drastically lowering memory and processing needs by translating the model's weights and activations from floating-point precision to lower-bit representations. Hardware acceleration is also thought to improve real-time processing. It entails making use of specialized hardware, such as TPUs, GPUs, and accelerators for deep learning applications. These hardware solutions are designed to efficiently perform large-scale matrix operations, allowing for real-time processing and significantly lowering inference time. To further simplify data handling and model execution, software optimizations are included, such as employing effective data pipelines and optimized libraries and frameworks. To improve throughput and lower delay during prediction, methods including batching and parallel processing are used. The proposed system is made to be scalable and deployable in clinical settings by taking into account these real-time implementation considerations. It ensures that the seizure prediction model can produce accurate and timely predictions, facilitating successful early intervention and enhancing patient care in real-world settings. In summary, the methodology provides a systematic way to predict seizures in epilepsy using LSTM-based RNNs. To improve clinical decision support systems for epilepsy management, prioritize careful model training and evaluation, reliable data collecting, preprocessing, and deep learning feature extraction. The methodology's technical precision and flexibility show potential for improving patient outcomes by enabling prompt seizure prediction and intervention. To promote wider clinical application, future research topics include improving model architectures, incorporating multimodal EEG data, and tackling practical implementation issues.

## IV. RESULTS AND DISCUSSION

The performance of the proposed RNN-based seizure prediction system against existing systems is assessed in the results and analysis section. Prediction accuracy, false alarm rate, and computing efficiency are important measures. The proposed system exhibits increased computing efficiency, decreased false alarm rates, and higher accuracy, indicating that real-time clinical applications could benefit from it.

TABLE I PREDICTION ACCURACY

| Metric | Existing System [7] | Existing System [8] | Proposed System |
|---|---|---|---|
| Sensitivity | 75.3 | 82.5 | 90.2 |
| Specificity | 80.1 | 84.7 | 88.9 |
| AUC -ROC | 78 | 86 | 93 |

Table I compares the accuracy of the proposed RNN-based seizure prediction system to that of existing systems [7] and [8]. Parameters tested include sensitivity, specificity, and the area under the receiver operating characteristic curve (AUC-ROC). With an AUC-ROC of 93, sensitivity of 90.2%, specificity of 88.9%, and enhanced performance, the proposed seizure onset prediction system exceeds the competition. The proposed system outperforms the existing system, which has a sensitivity of 75.3%, a specificity of 80.1%, and an AUC-ROC of 78. These findings show how well the RNN-based technique catches temporal dependencies in EEG data, leading to more precise and dependable seizure predictions.

TABLE II FALSE ALARM RATE BY PATIENT

| Patient ID | Existing System [7] | Existing System [8] | Proposed System |
|---|---|---|---|
| P1 | 16.0 | 12.1 | 7.3 |
| P2 | 14.5 | 11.0 | 6.4 |
| P3 | 157.7 | 10.8 | 6.7 |
| Average | 15.4 | 11.3 | 6.8 |

Table II presents a comparison of the false alarm rates for three different seizure prediction systems, P1, P2, and P3, as well as the Existing System [7] and [8] and the Proposed System. When compared to existing systems, the proposed system continuously exhibits the lowest false alarm rates (7.3%, 6.4%, and 6.7%). The existing system performs better than the proposed system (12.1%, 11.0%, and 10.8%), although it still has the greatest false alarm rates (16.0%, 14.5%, and 15.7%). The proposed system performs better in reducing erroneous predictions than the existing systems, with an average false alarm rate of 6.8%, which is far lower than the existing systems' 15.4% and 11.3%.

TABLE III COMPUTATIONAL EFFICIENCY

| Metric | Existing System [7] | Existing System [8] | Proposed System |
|---|---|---|---|
| Processing Time (ms) | 15 | 25 | 12 |
| Memory Usage (MB) | 50 | 120 | 45 |

Table III compares the computational efficiency of the proposed RNN-based seizure prediction system to the existing systems [7] and [8]. Two crucial requirements for real-time implementation are processing time and memory utilization, which are evaluated. The proposed system has a faster processing time of 12 milliseconds compared to existing systems that take 15 and 25 milliseconds, respectively. It shows that the proposed system's faster prediction-making skills are critical for timely responses. The proposed system consumes 45 MB of memory, compared to existing systems that use 50 MB and 120 MB, indicating efficient resource utilization. Because of these improvements in computational efficiency, the proposed system is now better suited for application in clinical settings requiring rapid and resource-efficient processing. In summary, the proposed RNN-based seizure prediction system performs noticeably better than the existing systems in terms of computing efficiency, false alarm rate, and prediction accuracy. It provides a better and more dependable tool for early seizure identification and prompt clinical intervention due to its improved sensitivity, specificity, AUC-ROC, fewer false alarms, and streamlined processing time and memory utilization. The proposed RNN-based seizure prediction system's results, which are shown in the discussion section, show significant improvements over the state-of-the-art techniques in terms of accuracy, false alarm rates, and computing efficiency. With a sensitivity of 90.2%, specificity of 88.9%, and an astounding AUC-ROC of 93, the system proved that it could successfully extract temporal dependencies in EEG data, which are essential for seizure prediction. It continuously showed lower false alarm rates (average of 6.8%), quicker processing times (12 ms), and lower memory consumption (45 MB) in comparison to the existing systems. The methodology can be used to improve clinical decision support for the management of epilepsy, facilitate early intervention, and possibly improve patient outcomes and quality of life. Its dependability in a range of therapeutic circumstances is attributed to its dynamic adjustments and capacity to react to the unique EEG patterns of specific patients. Furthermore, the technology is scalable and appropriate for wider clinical adoption due to its dependence on automated feature extraction from raw EEG data and deep learning. The proposed system has several advantages, such as strong performance measures that confirm its effectiveness in seizure prediction and computational efficiency, which is essential for real-time use in clinical situations. To further improve its usability and impact in the treatment of epilepsy, future research could concentrate on improving model designs, adding multimodal EEG data, and addressing real-world deployment issues.

## V. CONCLUSION

In conclusion, the proposed RNN-based examination signifies a noteworthy progression in seizure prediction for the management of epilepsy, providing superior accuracy, decreased false alarm rates, and increased computational efficiency in contrast to current techniques. The proposed system automatically adjusts to each patient's unique profile by directly extracting temporal relationships from raw EEG data using LSTM networks, which boosts prediction reliability. But even with these advantages, there are also a few drawbacks. First, the amount and diversity of the training dataset can affect the proposed system's performance. Secondly, additional optimization is required to ensure practical clinical deployment of real-time implementation challenges including hardware compatibility and latency reduction. Third, although the system functions well with EEG data, further research is necessary to see how well it works with other modalities as well as challenging clinical situations. Future development should concentrate on resolving these issues and enhancing the proposed system's functionalities. First off, generalizability and robustness can be increased by fine-tuning model architectures to handle bigger and more varied datasets. Second, including multimodal data sources, like extra physiological signals or patient-specific information, may improve forecast precision and increase applicability. Thirdly, for smooth incorporation into clinical workflows, real-time implementation tactics must be optimized through sophisticated hardware support and effective algorithmic advancements. By improving seizure prediction technologies, these initiatives will eventually improve patient outcomes and quality of life in the treatment of epilepsy.